\documentclass[12pt]{article}
\usepackage[utf8]{inputenc}
\usepackage{indentfirst} 
\usepackage{amsfonts,amssymb,amsmath}
\usepackage{bm}          
\usepackage{cite}
\usepackage{url}
\usepackage{enumitem}
\usepackage[usenames,dvipsnames]{xcolor}
\usepackage{todonotes}
\usepackage{multicol}
\usepackage{MnSymbol}
\usepackage{hyperref}
\usepackage{tikz-cd}
\usepackage[titletoc,title]{appendix}
\usepackage{tikz}
\usepackage{verbatim}
\usetikzlibrary{matrix}

\usetikzlibrary{arrows}

\AtBeginDocument{}%

\newcommand{\corurl}{red}
\newcommand{\corcite}{ForestGreen}
\newcommand{\corlink}{blue}

\hypersetup{linktocpage,colorlinks,urlcolor=\corurl,citecolor=\corcite,linkcolor=\corlink,
pdftitle={Boundary Hilbert spaces and trace operators},
pdfauthor={J.F. Barbero, J. Margalef-Bentabol, E.J.S. Villaseñor}}

\numberwithin{equation}{section}  
\begin{document}


\bibliographystyle{plain}

\title{Boundary Hilbert spaces and trace operators}

\author{
  {\small J. Fernando Barbero G.${}^{1,2}$, Benito A. Ju\'arez-Aubry${}^3$, Juan Margalef-Bentabol${}^{1,4}$,} \\[4mm]
  {\small  and Eduardo J.S. Villase\~nor${}^{2,4}$} \\[4mm]
  {\small\it ${}^1$Instituto de Estructura de la Materia, CSIC} \\[-0.2cm]
  {\small\it Serrano 123, 28006 Madrid, Spain}         \\[1mm]
  {\small\it ${}^2$Grupo de Teor\'{\i}as de Campos y F\'{\i}sica
             Estad\'{\i}stica}\\[-2mm]
  {\small\it Instituto Gregorio Mill\'an, Universidad Carlos III de
             Madrid}\\[-2mm]
  {\small\it Unidad Asociada al Instituto de Estructura de la Materia, CSIC}
             \\[-2mm]
  {\small\it Madrid, Spain}           \\[1mm]
  {\small\it ${}^3$Departamento de Gravitaci\'on y Teor\'{\i}a de Campos}\\[-2mm]
  {\small\it Instituto de Ciencias Nucleares, Universidad Nacional Aut\'onoma de M\'exico}\\[-2mm]
  {\small\it A. Postal 70-543, Mexico City 04510, Mexico}\\[1mm]
    {\small\it ${}^4$Grupo de Modelizaci\'on, Simulaci\'on Num\'erica
                   y Matem\'atica Industrial}  \\[-0.2cm]
  {\small\it Universidad Carlos III de Madrid} \\[-0.2cm]
  {\small\it Avda.\  de la Universidad 30, 28911 Legan\'es, Spain}            \\[1mm]
  {\protect\makebox[5in]{\quad}}  
  \\
}
\date{January 3, 2017}
\maketitle
\thispagestyle{empty}   

\begin{abstract}
We discuss the introduction of boundary Hilbert spaces for a class of physical systems for which it is not possible to factor their state spaces as tensor products of Hilbert spaces naturally associated to their boundaries and bulks respectively. In order to do this we make use of the so called \textit{trace operators} that play a relevant role in the analysis of PDE's in bounded regions. By taking advantage of these operators and some functorial aspects of the construction of Fock spaces, we will show how to obtain quantum dynamics at the boundaries defined in appropriate Hilbert spaces associated with them.
\end{abstract}


\clearpage

%
%
\section{Introduction}{\label{sec_intro}}

One of the fundamental tenets of the quantum mechanics of compound systems, a \textit{zeroth principle} according to some authors \cite{zurek}, is the fact that their state spaces are tensor products of Hilbert spaces associated with the constitutive subsystems (whose mutual interactions are accounted for by a Hamiltonian with non-zero matrix elements connecting the different factors). Although it may seem easy to understand and describe the parts of a compound system, a moment of reflection will suffice to realize that some care should be exercised. This is already true at the classical level.

We urge the reader to consider the following simple system. A piece of string of finite length attached at its ends to a pair of point masses. For the sake of the argument, let us suppose that the motion of the string and the masses is confined to one dimension. Despite the fact that the system seems to consist of three separate subsystems (the string and the two masses) it is easy to see that in a certain concrete sense this cannot be the case. This is so because its full configuration is completely determined by that of the string, i.e.\ the \textit{single} real function that associates to each point of the segment of length $\ell$, representing the unstretched string, its actual position. The natural continuity requirement for this function tells us what the positions of the masses at its ends are and, hence, it is redundant to specify them: \textit{There are no independent physical degrees of freedom associated with the point masses}.

A detailed and mathematically rigorous study of this example can be found in \cite{cuerdamasas}, where its Hamiltonian formulation, in an appropriate infinitely dimensional phase space, is thoroughly developed. An important ingredient of the construction given in \cite{cuerdamasas} is the use of appropriate functional spaces and mathematical objects such as discrete measures and Radon Nikodym derivatives. They encode in a convenient way the physical features of the system, for instance, the presence and effect of  the point masses can be neatly described by a measure with singular contributions. It is important to keep in mind that, despite the fact that no physical degrees of freedom are associated with the point masses, their presence and interaction with the whole system can (and must!) be taken into account by including appropriate terms in the Lagrangian. These allow us to incorporate, for example, the obvious effect of their inertia on the motion of the string.

If the dynamics of the system that we have discussed above is linear, its quantization can be conveniently carried out by Fock space methods (see, again, \cite{cuerdamasas}). As we showed in that paper, the Hilbert space of the system cannot be naturally written as the tensor product of factor Hilbert spaces associated, respectively, with the string and the masses. This is the quantum counterpart of the phenomenon that we have described in the two preceding paragraphs.

There are important physical situations in which we may want to concentrate only on the dynamics of the point particles attached to the string. In fact, the simple mechanical system consisting of two masses connected by a spring and discussed in basic physics courses is, precisely, of that type.  In the usual treatment of such a system one makes, more or less implicitly, the assumption that the internal degrees of freedom of the spring are irrelevant. After the preceding discussion we see that, both at the classical and quantum levels, the configuration of the whole system is such that no separate mathematical objects are needed to describe the dynamics of the masses. The following question then naturally comes up: how can we then get the simplified models where the string configurations seem to play no role and only the masses are relevant?

The purpose of this paper is to answer this question in the quantum setting. As we will show, the so called \textit{trace operators} ---not to be confused with the standard matrix or operator traces--- that allow us to associate boundary values to functions in appropriate functional spaces (of the Sobolev type, for instance) can be used for this purpose. The main idea is to consider these operators and ``exponentiate'' them by using Fock space techniques. This procedure will naturally lead us to introduce auxiliary Hilbert spaces where the dynamics of the point masses located at the ends of the string can be conveniently described.

In our opinion, the ideas put forward in this paper can be useful to understand in a rigorous way the physical implications that the presence of dynamical boundaries can have in many physical applications encompassing condensed matter systems and gravitational models. In particular, the possibility of associating gravitational degrees of freedom to horizons of different types (event horizons, isolated horizons\ldots) \cite{Ashtekar}, the holography paradigm \cite{tHooft}, and the AdS/CFT correspondence \cite{Maldacena,Witten} demand an understanding of the relatively subtle issues related to the definition of dynamics at boundaries and their interaction with the bulk.

The structure of the paper is the following: after this introduction we give a short review of previous results \cite{cuerdamasas} on the type of models considered in the paper. The main construction proposed in the paper is discussed in section \ref{sec_Fock}. Finally, our conclusions and some comments appear in section \ref{sec_conclusions}. Throughout the paper we will take $\hslash=1$.

%
%
\section{A short review of previous results}{\label{sec_review}}

In this paper we consider the dynamics of a class of systems with an infinite number of physical degrees of freedom and physical dynamical boundaries. In the past we have studied in detail \cite{cuerdamasas} an example of such a system: the longitudinal motion of an elastic string of finite length tied to two point masses located at its ends (themselves attached to springs of zero rest length). To make the present paper easier to read we give here some details on the most important features of that model.

The Lagrangian can be written in the form
\begin{equation}
L(Q,V)=\frac{1}{2}\langle V,V\rangle-\frac{1}{2}\langle Q', Q'\rangle-\frac{\widetilde{\omega}^2}{2}\langle Q,Q\rangle+\sum_{j\in\{0,1\}} \frac{\mu_j}{2}\Big(V(j)^2-\widetilde{\omega}_j^2Q(j)^2\Big)\,.
\label{lagrangian2}
\end{equation}
where the smooth $Q$ and $V$ functions denote the configuration variable (the field representing the deviation of the string point labelled by $x$ from its equilibrium position) and its velocity. In the previous expression $\langle \cdot ,\!\cdot \rangle$ denotes the standard $L^2(0,1)$ scalar product, and $\widetilde{\omega}$, $\widetilde{\omega}_j$ and $\mu_j$ ($j=0,1$) are constants related to the mass density of the string, its Young modulus, the spring constants and the masses of the particles attached at the ends of the string. Time derivatives are denoted by dots and spatial derivatives by primes.

The equations of motion are:
\begin{align}
\ddot{Q}(x,t)-Q''(x,t)+\widetilde{\omega}^2 Q(x,t)&=0\,,\quad x\in(0,1)\,,\label{eqmotion}\\
\mu_0 \ddot{Q}(0,t)-Q'(0,t)+\mu_0\widetilde{\omega}_0^2 Q(0,t)&=0\,, \label{eqmotion2}\\
\mu_1 \ddot{Q}(1,t)+Q'(1,t)+\mu_1\widetilde{\omega}_1^2 Q(1,t)&=0\,. \label{eqmotion3}
\end{align}

The time evolution of the deformation of the string is given by the 1+1 Klein-Gordon equation whereas the masses move under the combined force exerted by the springs and the string. The preceding equations are not standard because \eqref{eqmotion2} and \eqref{eqmotion3} are not ordinary boundary conditions as they involve second order time derivatives. This feature qualitatively changes the type of eigenvalue problem that has to be solved to get the normal modes and characteristic frequencies. In fact, the relevant eigenvalue equations \textit{are not} of the Sturm-Liouville type and, as a consequence, the theorems that guarantee that the normal modes form a complete set cannot be applied.

As we show in \cite{cuerdamasas} a neat way to circumvent this issue is to introduce a new \textit{measure} on the string with singular contributions at its ends. With its help it is possible to define a self-adjoint oparator $\Delta$ and describe the dynamics of the system with a wave equation $-\ddot{Q}+\Delta Q=0$ subject to boundary conditions of the Robin type (written in terms of the relevant Radon-Nikodym derivative). In a sense this idea is similar to the introduction of non-trivial metrics characteristic of QFT in curved spacetimes \cite{Wald}. The role of the new measure is crucial to understand the one particle Hilbert space of the model and the  Fock space $\mathcal{H}$ built from it. The non-factorizability of $\mathcal{H}$ as a tensor product of Hilbert spaces naturally associated to the string and the masses (the main result of \cite{cuerdamasas}) can be traced back to the specific features of the new measure. Although in the following we will not have to rely of the details of the string-masses model it is useful to keep it in mind. For most of the arguments that we present here it is enough to consider a linear physical system with boundaries and use some generic features of the abstract wave equation.

%
%
\section{Fock space quantization}{\label{sec_Fock}}

\subsection{Fock spaces}{\label{sec_Fock_spaces}}

Let us consider a generic linear physical system with an infinite number of ``field-like'' degrees of freedom. The dynamics of such a system will be generically given by an abstract wave equation (see \cite{Marsden,Hamiltonian}) of the type
\begin{equation*}
  \ddot{\varphi}-\Delta \varphi=0\,,
\end{equation*}
where $\varphi(t,\cdot)$ is defined on a certain smooth manifold $\Sigma$ with boundary $\partial\Sigma$ and, for the sake of the arguments presented in the paper, will be taken as a curve in a real Hilbert space where the operator $\Delta$ is self-adjoint. This type of models is rather general because it describes not only the evolution of fields in curved stationary backgrounds (in this case $\Delta$ is the Laplace-Beltrami operator) but also other models where the interaction of objects with different dimensions is incorporated by using singular measures \cite{cuerdamasas}.

In order to simplify our arguments $\Sigma$ will be simply a closed interval $[\alpha,\beta]$ of the real line. It is important to mention at this point that, despite the apparent simplicity of the setting that we have just laid out, it is general enough to contain the class of compound models that we intend to study here. As discussed at length in \cite{cuerdamasas}, by introducing a non-trivial measure with singular contributions at the boundary of $\Sigma$ and using it to define the Hilbert space of solutions to the equations of motion, we can take into account the dynamics of the string-masses system.

From the real Hilbert space of solutions to the field equations it is straightforward to build the complex 1-particle Hilbert space $\mathfrak{h}$ by introducing a notion of positive frequency and taking only the (complex) positive frequency solutions. The evolution in $\mathfrak{h}$ is defined by a 1-particle classical Hamiltonian $h$: a self-adjoint operator of the general form $\sqrt{-\Delta}$ where $\Delta$ can be thought of as a generalized Laplacian. The unitary evolution operator in $\mathfrak{h}$ is given by $\exp(-it h)$. Given initial data $\varphi_0$ and $\dot{\varphi}_0$ for the field it is possible to build a complex vector $v_0\in \mathfrak{h}$ so that the classical evolution of the system is encoded in the curve
\begin{equation*}
  v_t:=\exp(-i t h)v_0\,.
\end{equation*}
The quantization of the system is carried out by defining a Fock space with the help of the 1-particle Hilbert space just introduced (see, for instance,\cite{Bratteli} and \cite{Wald}). Here we consider the symmetric Fock space $\mathcal{H}:=\Gamma(\mathfrak{h})$ as we are dealing with bosonic objects. The unitary evolution operator $U(t):=\Gamma(\exp(-ith))$ that defines the dynamics of the system in Fock space $\mathcal{H}$ acts on its elements in the following way
\begin{align*}
  & U(t)\Omega=\Omega\,,  \\
  & U(t)(v_1\circ v_2\circ\cdots\circ v_n)=\exp(-iht)v_1\circ\exp(-iht)v_2\circ\cdots\circ \exp(-iht)v_n\,.
\end{align*}
In the preceding expression $\Omega$ denotes the so called ``vacuum state'' and the symmetrized tensor product is denoted by $\circ$. In order to write down the expression for the self-adjoint Hamiltonian operator
\[
H:=\mathrm{d}\Gamma(h)=i\left.\frac{\mathrm{d}\,}{\mathrm{d}t}\right|_{t=0} U(t)
\] in the Fock space $\mathcal{H}$, it suffices to differentiate the preceding expression at $t=0$. This way we get
\begin{align*}
  & H\Omega=0\,,\\
  & H (v_1\circ v_2\circ\cdots\circ v_n)=hv_1\circ v_2\circ\cdots\circ v_n+v_1\circ hv_2\circ\cdots\circ v_n+\cdots+v_1\circ \cdots\circ v_{n-1}\circ hv_n\,.
\end{align*}

\subsection{Coherent states}{\label{sec_coherent_states}}

The so called coherent states play a central role in the discussion of linear quantum systems. They are convenient generalized bases of states that provide us with a useful connection between the dynamics of the classical system and its quantum counterpart (obtained through the construction that we have just sketched). This is so because, in a definite sense, the evolution of the coherent states just mimics that of the classical states used to define them. Indeed, let us take $v\in\mathfrak{h}$ (remember that such a vector encodes the initial data information both for ``position'' and ``velocity'') and build the  [non-normalized] coherent state $\varepsilon(v)$ in the symmetrized Fock space $\mathcal{H}$
\begin{equation*}
  \varepsilon(v):=\sum_{n=0}^\infty\frac{v^{\circ n}}{\sqrt{n!}}=\Omega+v+\frac{v\circ v}{\sqrt{2!}}+ \frac{v\circ v\circ v}{\sqrt{3!}}+\cdots\,.
\end{equation*}
Its (unitary) time evolution is given by
\begin{equation*}
  U(t)\varepsilon(v)=\sum_{n=0}^\infty\frac{(e^{-ith}v)^{\circ n}}{\sqrt{n!}}=\varepsilon(\exp(-iht)v)=\varepsilon(v_t)\,.
\end{equation*}
It is, hence, clear the sense in which the quantum evolution mimics the classical one: the evolved coherent state is labeled by the vector that encodes the classical solution of the field configuration whose initial data are defined by $v$.

On these coherent states the Hamiltonian acts in the following simple way
\begin{equation}
  H\varepsilon(v)=a^*(hv)\varepsilon(v)\,,\label{He}
\end{equation}
where, given $u\in \mathcal{H}$, we define the usual (bosonic) annihilation and creation operators $a(u)$ and $a^*(u)$ as
\begin{align*}
  &a(u)\Omega=0\,,\\
  &a(u)(v_1\circ v_2\circ\cdots\circ v_n)=\frac{\langle u,v_1\rangle}{\sqrt{n}} v_2\circ\dots \circ v_n+\cdots+\frac{\langle u,v_n\rangle}{\sqrt{n}}v_1\circ\dots \circ v_{n-1}\,,\\
  &a^*(u)\Omega=u\,,\\
  &a^*(u)(v_1\circ v_2\circ\cdots\circ v_n)=\sqrt{n+1}\ u\circ v_1\circ\cdots\circ v_n\,.
\end{align*}
In the preceding expressions $\langle\cdot,\cdot\rangle$ denotes the scalar product in the 1-particle Hilbert space $\mathfrak{h}$. Notice that $a(u) v^{\circ n}=\sqrt{n}\langle u,v\rangle v^{\circ(n-1)}$ and hence
\begin{equation}
a(u)\varepsilon(v)=\langle u,v\rangle \varepsilon(v)\,.\label{ae}
\end{equation}
Using \eqref{He} and \eqref{ae} (when $v\neq 0$) we can write
\[
H\varepsilon(v)=\frac{a^*(hv)a(v)}{||v||^2}\varepsilon(v)\,.
\]
It is interesting to point out that the Fock space $\mathcal{H}=\Gamma(\mathbb{C})$ associated with $\mathbb{C}$ is isomorphic to $L^2(\mathbb{R})$ (or $\ell^2(\mathbb{N})$ for that matter).
In this case the one particle Hamiltonian is simply a multiplicative operator $hv=\omega v$ and, hence,
\[
H\varepsilon(v)=\omega N \varepsilon(v)
\]
where $N=a^*(1)a(1)$ is the number operator written in terms of the usual annihilation and creation operators $a(1)$ and $a^*(1)$ in $\Gamma(\mathbb{C})$.

\subsection{Second quantization functor}{\label{sec_functor}}

The main instrument that we use in the paper is the following simple idea: Given a linear map between two Hilbert spaces $T:\mathfrak{h}_I\rightarrow \mathfrak{h}_F$ it is possible to define in a natural (functorial) way an operator between the associated Fock spaces $\mathcal{H}_i =\Gamma(\mathfrak{h}_i)$ by giving its action on the (overcomplete set of) coherent states. In detail
\begin{center}
\begin{tikzpicture}
  \matrix (m) [matrix of math nodes,row sep=3em,column sep=4em,minimum width=2em]
  {
    \mathcal{H}_I & \mathcal{H}_F \\
   \mathfrak{h}_I  & \mathfrak{h}_F \\};
  \path[-stealth]
    (m-2-1) edge node [left] {$\Gamma$} (m-1-1)
    (m-2-1) edge node [below] {$T$} (m-2-2)
    (m-1-1.east|-m-1-2) edge node [above] {$\Gamma(T)$}
            (m-1-2)
    (m-2-2) edge node [right] {$\Gamma$} (m-1-2)
            ;
\end{tikzpicture}
\end{center}
The action of $\Gamma(T)$ on the coherent states is
\begin{align}
\Gamma(T)\varepsilon_I(v):=\varepsilon_F(Tv)\,.\label{lift2}
\end{align}
The operator $\Gamma(T)$ is customarily known as the \textit{second quantization} of $T$. In practice it suffices to have $T$ defined in a dense subspace of $\mathfrak{h}_I$.

Before explaining how this idea works in the case of interest for us, let us first discuss the following simple example. Consider a linear system with a discrete orthonormal Hilbert basis of ``normal modes'' $\varphi_n^+\in\mathfrak{h}$ (vectors in the 1-particle Hilbert space). For a concrete index $n$ let us define the coordinate map
$$
\pi_n:\mathfrak{h}\rightarrow \mathbb{C}\,,\quad v=\sum_kv_k\varphi_k^+\mapsto\pi_n(v)=v_n\,.
$$
The second quantization of $\pi_n$ is given by
\begin{align*}
&\Gamma(\pi_n):\mathcal{H}\rightarrow \mathcal{H}_n:=\Gamma(\mathbb{C})\,,\quad \varepsilon(v)\mapsto\varepsilon_n(v_n)\,,
\end{align*}
where $\varepsilon_n(z)$ is the (standard) coherent state for a one-dimensional harmonic oscillator labeled by $z\in\mathbb{C}$.

Let us study now the action of $ \Gamma(\pi_n)$ on the curve $U(t)\varepsilon(v)$ given by the unitary evolution of the coherent state $\varepsilon(v)\in \mathcal{H}$:
\begin{equation*}
  \Gamma(\pi_n)U(t)\varepsilon(v)=\Gamma(\pi_n)\varepsilon(v_t)=\varepsilon_{n}(\pi_n(v_t))=\varepsilon_n (e^{-i\omega_nt}v_n)\,,
\end{equation*}
where $\omega_n$ denotes the frequency of the $n^{\mathrm{th}}$ normal mode. It is straightforward now to show that the time-dependent vectors $\psi_t:=\varepsilon_n (e^{-i\omega_nt}v_n)$ satisfy, irrespectively of $v\in \mathfrak{h}$, the Schr\"odinger equation
\begin{equation}\label{coordinates}
  i\dot{\psi}_t=\omega_nN_n \psi_t\,,
\end{equation}
where $N_n$ is the number operator in $\mathcal{H}_n$. Indeed, using \eqref{He} and \eqref{ae} we have
\begin{align*}
  i\dot{\psi}_t&=a_n^*(\omega_ne^{-i\omega_n t}v_n)\varepsilon_n(e^{-i\omega_nt}v_n)=\omega_ne^{-i\omega_nt}v_na_n^*(1)\varepsilon_n(e^{-i\omega_nt}v_n) \\
                     &=\frac{\omega_ne^{-i\omega_nt}v_n}{e^{-i\omega_nt}v_n}a_n^*(1)a_n(1)\varepsilon_n(e^{-i\omega_n t}v_n)=\omega_nN_n\psi_t\,.
\end{align*}
where $a_n(1)$ and $a^*_n(1)$ are the annihilation and creation operators in $\mathcal{H}_n$.

\subsection{Inclusion of boundaries: trace operators}{\label{Trace operators}}

The so called trace operator plays a central role in the study of PDE's. In a nutshell it serves the purpose of defining the boundary values of the elements of the functional spaces (typically of the Sobolev type) used to discuss PDE's in bounded regions (see \cite{Brezis} for a pedagogical exposition). In order to understand the relevance of such an operator it helps to keep in mind that the elements of $L^{p}(\mathbb{R})$ spaces are classes of functions for which there is no way to assign a value to an specific point $x\in \mathbb{R}$. One of the virtues of the Sobolev spaces used to discuss PDE's in bounded regions is the possibility of associating boundary values for their elements (by introducing appropriate functional spaces at the boundaries). The operator $\gamma:H^1(\Sigma)\rightarrow L^2(\partial\Sigma)$ that performs the trick is known in the PDE literature as the \textit{trace} (not to be confused with the trace of a linear operator). In the simplest of cases, that of a closed interval $\Sigma=[\alpha,\beta]$ of the real line, the trace actually associates real numbers to both $\alpha$ and $\beta$, in fact
\begin{equation*}
  \gamma:H^1([\alpha,\beta])\rightarrow \mathbb{C}^2\,, \quad f\mapsto \gamma(f):=(\gamma_\alpha f,\gamma_\beta f)\,,
\end{equation*}
where $H^1([\alpha,\beta])$ denotes the first Sobolev space on $[\alpha,\beta]$. It is interesting to point out that in this particular example every element of $H^1([\alpha,\beta])$ has a continuous representative in $C([\alpha,\beta])$ (see \cite{Brezis}) and the trace is just given by its values at $\alpha$ and $\beta$.

In more complicated circumstances the smoothness properties of the boundary play a significant role in the definition of the trace and the images of the trace operators take values in quite non-trivial functional spaces. The interested reader is referred again to \cite{Brezis}.

\subsection{Second quantization of the trace operator}{\label{Trace operators_lift}}

In the case that we have discussed above the one particle Hilbert space $\mathfrak{h}$ is of the Sobolev type \cite{cuerdamasas} and, hence, an appropriate trace operator
$$
\gamma:\mathfrak{h}\rightarrow \mathbb{C}^2: v\mapsto \gamma(v)=(\gamma_\alpha v,\gamma_\beta v)
$$
exists. Notice that the boundary of $\Sigma$ consists of two points. We can, hence, define two ``partial'' traces $\gamma_\alpha$ and $\gamma_\beta$  associated with $\alpha$ and $\beta$. To simplify the following discussion we concentrate on one of them. Let us consider the second quantization of, say $\gamma_\alpha:\mathfrak{h}\rightarrow \mathbb{C}$,  and use the construction outlined above
\begin{align*}
  & \Gamma(\gamma_\alpha):\mathcal{H}\to \mathcal{H}_\alpha:=\Gamma(\mathbb{C})\,,\quad\Gamma(\gamma_\alpha)\varepsilon(v)=\varepsilon_\alpha(\gamma_\alpha v)\,,
\end{align*}
where, here, $\varepsilon_\alpha$ denotes the coherent states on the boundary Hilbert space $\mathcal{H}_\alpha$ associated with the point $\alpha$.

We can study now the dynamics on the boundary Hilbert space by following the same approach that we employed for the operators $\pi_n$ considered above. Let us take a vector $v$ in the 1-particle Hilbert space $\mathfrak{h}$, build the coherent state $\varepsilon(v)$ and study the action of the second quantization of $\gamma_\alpha$ on the curve $U(t)\varepsilon(v)$ given by the evolution of the coherent state $\varepsilon(v)\in \mathcal{H}$
\begin{equation*}
  \Gamma(\gamma_\alpha)U(t)\varepsilon(v)=\varepsilon_\alpha(\gamma_\alpha v_t)\,,\quad\gamma_\alpha v_t\in \mathbb{C}\,.
\end{equation*}
It is easy to see now that the time-dependent vectors $\displaystyle \psi_t^\alpha:=\varepsilon_\alpha(\gamma_\alpha v_t)\in \mathcal{H}_\alpha$ are solutions to the Schr\"{o}dinger-like equation
\begin{equation*}
  \dot{\psi}_t^\alpha=-i\omega_\alpha(v_t)N_\alpha \psi_t^\alpha\,,\quad \omega_\alpha(v_t):=\frac{\gamma_\alpha h v_t}{\gamma_\alpha v_t}\in \mathbb{C}\,,
\end{equation*}
where $N_\alpha$ denotes the number operator in $\mathcal{H}_\alpha$. Notice that the operators $\omega_\alpha(v_t)N_\alpha$ are not necessarily self-adjoint because, generically, $\omega_\alpha(v_t)\not\in \mathbb{R}$. The argument is very similar to the one leading to \eqref{coordinates}
\begin{align*}
  i\dot{\psi}_t^\alpha&=a_\alpha^*(\gamma_\alpha h v_t)\varepsilon_\alpha(\gamma_\alpha v_t)=(\gamma_\alpha h v_t)a^*_\alpha(1)\varepsilon_\alpha(\gamma_\alpha v_t)\\
                     &=\frac{\gamma_\alpha h v_t}{\gamma_\alpha v_t}a_\alpha^*(1)a_\alpha(1)\varepsilon_\alpha(\gamma_\alpha v_t)=\omega_\alpha(v_t)N_\alpha\psi_t^\alpha\,.
\end{align*}
Several comments are in order now. First, notice that the evolution of ${\psi}_t^\alpha$ is determined by a non self-adjoint ``Hamiltonian'' $\omega_\alpha(v_t)N_\alpha$ that depends on the classical trajectory through the vector $v_t\in\mathfrak{h}$. This is different from the standard situation in which the Hamiltonian is simply an operator defined on an appropriate subspace of the Hilbert space for the system. An important consequence of this is that the norm of $\psi_t^\alpha$ changes with time. Indeed, the scalar product of two boundary coherent states is
\begin{equation*}
\langle \varepsilon_\alpha(\gamma_\alpha u_t), \varepsilon_\alpha(\gamma_\alpha v_t) \rangle=\exp({\overline{\gamma_\alpha u_t}\cdot \gamma_\alpha v_t})\,,
\end{equation*}
and, applying \eqref{ae}, we see that it evolves in time according to the equation
\begin{equation*}
\frac{\mathrm{d}}{\mathrm{d}t}\langle \varepsilon_\alpha(\gamma_\alpha u_t), \varepsilon_\alpha(\gamma_\alpha v_t) \rangle=i(\overline{\gamma_\alpha h u_t}\cdot\gamma_\alpha v_t- \overline{\gamma_\alpha u_t}\cdot\gamma_\alpha h v_t)\langle \varepsilon_\alpha(\gamma_\alpha u_t), \varepsilon_\alpha(\gamma_\alpha v_t) \rangle\,.
\end{equation*}
Thus, if $\gamma_\alpha v_t\neq0$, the previous definition of $\omega(v_t)$ leads to
\begin{equation*}
\frac{\mathrm{d}}{\mathrm{d}t}\|\varepsilon_\alpha(\gamma_\alpha v_t)\|^2=2\mathrm{Im}\left(\omega_\alpha(v_t)\right)|\gamma_\alpha v_t|^2\|\varepsilon_\alpha(\gamma_\alpha v_t)\|^2\,\quad \mathrm{with}\quad \|\varepsilon_\alpha(\gamma_\alpha v_t)\|^2=e^{|\gamma_\alpha v_t|^2}
\end{equation*}
This expression is, generically, different from zero as can be easily seen by taking a sum of normal modes $v=\varphi^+_1+\varphi^+_2$ and computing
\begin{equation*}
\omega_\alpha(v_t)=\frac{\omega_1\varphi^+_1(t,\alpha)+\omega_2\varphi^+_2(t,\alpha)}{\varphi^+_1(t,\alpha)+\varphi^+_2(t,\alpha)}\,,
\end{equation*}
which has an oscillating imaginary part on account of the different time-dependent phases of the numerator and denominator.

Although the present situation is somewhat unfamiliar, a similar phenomenon happens already at the classical level, because in that case it is also impossible, in general, to describe the dynamics of the masses attached at the ends of the string by using a single Hamiltonian function (as can be seen by solving the equations of motion for the system). At any rate, by adopting a physical perspective, it is not difficult to identify situations in which an approximate Hamiltonian can be written. Consider, for instance, ``equilibrium initial data'' i.e.\ a starting point in which the string is stretched by pulling the masses, waiting until it reaches equilibrium and then released. In such circumstances the masses should move  ---at least for some time--- as if they were effectively connected by a spring with no internal degrees of freedom, a situation that is obviously described by a simple harmonic oscillator Hamiltonian.

An important comment is in order. In the process of defining a ``boundary Hilbert space'' (remember that it must be introduced by hand because it does not appear as a factor in a tensor product decomposition of the full Hilbert space of the system) we are not just particularizing the values of the quantum fields to the points of the boundary. Strictly speaking the quantum fields are not defined at spacetime points (this is often expressed in the literature by saying that \textit{quantum fields are distributional}) but, rather, are labeled by particular vectors in the 1-particle Hilbert space
\begin{align*}
  \Phi(v) & :=\frac{1}{\sqrt{2}}\big(a(v)+a^*(v)\big)\,, \\
  \Pi(v) & :=\frac{1}{i\sqrt{2}}\big(a(v)-a^*(v)\big)\,.
\end{align*}
Notice that these objects are related because $\Phi(i v)=\Pi(v)$. The expectation value of the field $\Phi(u)$ in a coherent state $\varepsilon(v_t)$ is simply given by $\langle \Phi(u) \rangle_{\varepsilon (v_t)}=\sqrt{2}\mathrm{Re}\langle u,v_t \rangle$.  It is clear that all  the information contained in the classical evolution can be retrieved from the previous expectation value by a suitable choice of $u$. Notice also that the full evolution of the system is encoded in the expectation value of the annihilation operator as $\langle a(u)\rangle_{\varepsilon(v_t)}=\langle u, v_t\rangle$.

In the Hilbert space $\Gamma(\mathbb{C})$ we can define position and momentum operators similar to the field operators  $\Phi(v)$, $\Pi(v)$ in $\Gamma(\mathfrak{h})$, and postulate that they provide measurements on the position of the particles at the end of the string $x=\alpha$
\begin{align}
\Phi_\alpha(z) & :=\frac{1}{\sqrt{2}}\big(a_\alpha(z)+a_\alpha^*(z)\big)\,, \label{positionboundary}\\
\Pi_\alpha(z) & :=\frac{1}{i\sqrt{2}}\big(a_\alpha(z)-a_\alpha^*(z)\big)\,,\label{momentumboundary}
\end{align}
for $z\in \mathbb{C}$.  Notice that $X_\alpha:=\Phi_\alpha(1)$ and $P_\alpha:=\Pi_\alpha(1)$ are the standard position and momentum operator in $L^2(\mathbb{R})$ defined in terms of the creation and annihilation operators. It is very important to realize that in the present case we have $\langle X_\alpha \rangle_{\varepsilon_\alpha (\gamma_\alpha v_t)}=\sqrt{2}\mathrm{Re} (\gamma_\alpha v_t) $ so it may be sensible to postulate that measurements of $X_\alpha$ can indeed be interpreted as giving information about the position of the particle at $\alpha$ in the instant $t$.

The introduction of the partial trace associated with the other boundary $\beta$ follows exactly the same steps. By tensoring the two Hilbert spaces $\mathcal{H}_\alpha\otimes\mathcal{H}_\beta$ we can get the quantum dynamics in the whole boundary of $\Sigma$. In particular the coherent states at any instant of time have the form $\varepsilon_\alpha(\gamma_\alpha v_t)\otimes \varepsilon_\beta(\gamma_\beta v_t)$.

\bigskip

%
%
\section{Conclusions and comments}{\label{sec_conclusions}}

The main result of the paper is a proposal to introduce quantum dynamics at boundaries for systems whose Hilbert space cannot be naturally factored in terms of Hilbert spaces associated with the bulk and the boundary. Although these systems are not often discussed in the literature, they appear in the most simple examples where objects of different dimensions, such as strings and masses, are coupled (see  \cite{cuerdamasas}). They also play an important role in the study of dissipative phenomena in quantum mechanics (as discussed in \cite{Yurke1,Yurke2} and also in \cite{CL}).

In the opinion of the authors, systems of this type could play a relevant role in  condensed matter physics, general relativity in the presence of boundaries and the discussion of particle detectors in QFT in curved spacetimes. We would like to make some comments regarding this last issue. The standard treatment of detectors in the usual framework relies on the hypothesis that joint field-detector states can be written, at least in some regime, as tensor products. This is, for instance, one of the elements in the construction of the Unruh-Wald detectors \cite{UW} where a two level system is linearly coupled to a field and the detection of particles is carried out by observing excitations from the ground state to the (single) excited state of the detector. In the example that we have discussed  in \cite{cuerdamasas} and here it is not straightforward to see particles in this way because it is not possible to talk about a boundary Hamiltonian. Notice, however, that we do have boundary observables such as \eqref{positionboundary} (a position operator)  that can be measured in principle and have expectation values with a clear physical interpretation because $\langle X_\alpha \rangle_{\varepsilon_\alpha (\gamma_\alpha v_t)}=\sqrt{2}\mathrm{Re}(\gamma_\alpha v_t)$. This works despite the fact that we cannot talk of unitary evolution at the boundary for $\psi^\alpha_t$ (for the same reason the existence and relation between the Heisenberg and Schr\"{o}dinger pictures at the boundary is problematic).

As we have shown there are natural ways to lift operators acting between 1-particle Hilbert spaces to the Fock spaces built with their help. A very useful operator in the present context is the so called \textit{trace}. This kind of operator plays an important role in the rigorous mathematical study of PDE's and leads to a convenient definition of what a boundary value is in appropriate functional spaces. We use the second quantization process to define ``quantum boundary values'' of the field on auxiliary Hilbert spaces associated with the boundary of the system discussed in \cite{cuerdamasas}. By concentrating on the dynamics of appropriate coherent states, it is easy to see that the procedure that we propose achieves its purpose as the dynamics at the boundary (the motion of the masses attached at the ends of the string) is the classical one.

We would like to emphasize here that the construction that we have presented above is, actually, very natural from a mathematical point of view as it makes use of the simplest and more natural objects available to rigorously define a physical Hilbert space at the boundary. To what extent the structures that we have introduced are relevant from a physical perspective remains to be seen, in any case they seem capable of providing us with means to define idealized operational measurement procedures involving boundaries.

One last comment. In our opinion the use of trace operators, natural for coupled systems like the one that we have discussed in the paper, offers the interesting possibility of introducing a sort of non-unitary dynamics that may illuminate issues related to the collapse of the wave function or the surprising quantum behavior of gravitational systems in the presence of boundaries. We plan to look at these problems in the near future.

%
%
\section*{Acknowledgments}

This work has been supported by the Spanish MINECO research grant FIS2014-57387-C3-3-P.  Benito A Ju\'arez-Aubry is supported by CONACYT, project 101712. Juan Margalef-Bentabol is supported by a ``la Caixa'' fellowship and a Residencia de Estudiantes (MINECO) fellowship.

%
%

\end{document}